\documentclass[sigconf]{acmart}

\AtBeginDocument{%
  \providecommand\BibTeX{{%
    \normalfont B\kern-0.5em{\scshape i\kern-0.25em b}\kern-0.8em\TeX}}}

\copyrightyear{2023}
\acmYear{2023}
\setcopyright{rightsretained}
\acmConference[CHI EA '23]{Extended Abstracts of the 2023 CHI
Conference on Human Factors in Computing Systems}{April 23--28,
2023}{Hamburg, Germany}
\acmBooktitle{Extended Abstracts of the 2023 CHI Conference on Human
Factors in Computing Systems (CHI EA '23), April 23--28, 2023,
Hamburg, Germany}
\acmDOI{10.1145/3544549.3573849}
\acmISBN{978-1-4503-9422-2/23/04}

\begin{document}

\title{AI for human assessment: What do professional assessors need?}

\author{Riku Arakawa}
\orcid{0000-0001-7868-4754}
\affiliation{%
  \institution{Carnegie Mellon University}
  \city{Pittsburgh}
  \country{USA}
   \and
  \institution{ACES, Inc.}
  \city{Tokyo}
  \country{Japan}
}
\email{rarakawa@cs.cmu.edu}
\authornote{These authors contributed equally and are ordered alphabetically.}

\author{Hiromu Yakura}
\orcid{0000-0002-2558-735X}
\affiliation{
    \institution{University of Tsukuba / National Institute of Advanced Industrial Science and Technology (AIST)}
    \city{Tsukuba}
    \country{Japan}
}
\email{hiromu.yakura@aist.go.jp}
\authornotemark[1]

\renewcommand{\shortauthors}{Arakawa and Yakura}

\newcommand{\eg}{\textit{e.g.},~}
\newcommand{\ie}{\textit{i.e.},~}

\begin{abstract}
Recent organizations have started to adopt AI-based decision support tools to optimize human resource development practices, while facing various challenges of using AIs in highly contextual and sensitive domains.
We present our case study that aims to help professional assessors make decisions in human assessment, in which they conduct interviews with assessees and evaluate their suitability for certain job roles.
Our workshop with two industrial assessors elucidated troubles they face (\ie maintaining stable and non-subjective observation of assessees' behaviors) and derived requirements of AI systems (\ie extracting their nonverbal cues from interview videos in an interpretable manner).
In response, we employed an unsupervised anomaly detection algorithm using multimodal behavioral features such as facial keypoints, body and head pose, and gaze.
The algorithm extracts outlier scenes from the video based on behavioral features as well as informing which feature contributes to the outlierness.
We first evaluated how the assessors would perceive the extracted cues and discovered that the algorithm is useful in suggesting scenes to which assessors would pay attention, thanks to its interpretability.
Then, we developed an interface prototype incorporating the algorithm and had six assessors use it for their actual assessment.
Their comments revealed the effectiveness of introducing unsupervised anomaly detection to enhance their feeling of confidence and objectivity of the assessment along with potential use scenarios of such AI-based systems in human assessment.
Our approach, which builds on top of the idea of separating observation and interpretation in human-AI collaboration, will facilitate human decision making in highly contextual domains, such as human assessment, while keeping their trust in the system.
\end{abstract}





\newcommand{\tabref}[1]{Table~\ref{#1}}
\newcommand{\figref}[1]{Figure~\ref{#1}}
\newcommand{\secref}[1]{Section~\ref{#1}}
\newcommand{\eqnref}[1]{Equation~\ref{#1}}
\newcommand{\appref}[1]{Appendix~\ref{#1}}

\maketitle

\section{Introduction}
\label{sec:intro}

Human assessment is a process that aims to evaluate and make decisions on candidates regarding their suitability for certain types of employment \cite{Sartori2020Psychological}.
It originated from the selection of military officers during World War II \cite{ballantyne2004assessment} and now plays an important role in human resource development, especially for management jobs \cite{Bray1966TheAssessment}, where the skills of candidates as managers in their organizations are examined.
For example, in Germany, 73.4\% of the DAX-100 (German stock index) companies employ human assessment to evaluate their employees \cite{Obermann2008}.
The assessment process involves multiple methods such as job-related simulations, interviews, and psychological tests.
Among them, interviews are commonly used and viewed as a reliable method \cite{Roberts2014TheValidity} in which professional assessors conduct short interviews with candidate assessees.

However, the current workflow of interviews in human assessment contains some troubles.
For example, assessors often need to review the interviews that are video-recorded to manually check them in detail before making final decisions, which is time-consuming and mentally-demanding.
In addition, it is pointed out that, in such situations, errors tend to occur due to assessors' subjectivity \cite{Roberts2014TheValidity}, which hinders fair assessment as well as effective feedback to assessees.

Given the development of techniques for human behavior analysis during a conversation \cite{DBLP:books/cu/p/RudovicNP17, DBLP:conf/chi/ArakawaY19}, we speculated that it is possible to develop computational support to facilitate the assessment process.
However, it remains to be explored what the appropriate systems are for achieving such human-AI collaboration in human assessment.
For example, as Arrieta et al. \cite{DBLP:journals/inffus/ArrietaRSBTBGGM20} discussed, it is obviously not a good idea to develop a black-box prediction system that simply outputs a score for each candidate based on the recorded video of their interview.
This is because it would be hard for assessors to rely on the system's output without any explanation, especially in a field like human assessment, where they need to make a highly complicated and sensitive decision \cite{Glikson2020}.
Giermindl et al. \cite{doi:10.1080/0960085X.2021.1927213} alerted the \textit{dark side} of introducing AI-based decision making in human resource development, as it would lead to a lack of transparency and accountability and a reduction of employees' autonomy.
In other words, we need to design a human-AI relationship where assessors' trust in AI systems can be nurtured and to develop a plausible system that helps them through the decision-making process.

In this paper, we report on our story of designing computational support for human assessment through a series of studies.
The workshop we conducted with professional assessors discovered anti-patterns in automating human assessment as well as a potential area to which AI can contribute under the assumption of its interpretability.
Given that, we anticipated that it might be possible to monitor assessees' nonverbal behaviors by using computer vision techniques for extracting informative cues from interview videos, which would help assessors' decision making.
As such, we introduced an interpretable unsupervised anomaly detection with multimodal behavioral features as input to achieve the design of \textit{separating observation and interpretation}.
Our algorithm evaluation suggested that the algorithm can capture important behavioral cues as well as reconfirming the interpretability of the AI's output.
Our subsequent system evaluation revealed that assessors who used our prototype for their assessment felt that their assessment quality was improved with enhanced confidence and objectivity thanks to the AI's observation.
It also suggested room for improvement and potential use scenarios of the employed AI technique in human assessment.
We believe our case study will serve as an example of achieving human-AI collaboration in an area that involves highly human context and sensitive influences.

\section{Workshop: Derivation of System Requirements}
\label{sec:requirements}

To identify the requirements for the supporting tool in human assessment, we first conducted a workshop with assessors from a Japanese human-assessment company.
This company conducts approximately 2,000 assessment sessions annually and has a history of more than 25 years, mainly to check the suitability as company managers of the employees at their client enterprises.
The workshop was organized informally and went on to cover various aspects of developing such a system through conversation.
We involved two proficient assessors, who regularly manage and educate other assessors as well as conducting assessments.
The workshop lasted approximately three hours in total.

\subsection{How the assessment is conducted}
\label{sec:requirements-how}

First, we ask the assessors about the overview of their assessment routine to get ourselves familiarized with the process.
As follows, they described the details of the process consisting of two sessions: an interview phase and a review phase.

In the interview phase, an assessor plays a certain role and seeks to evaluate how their assessee behaves in the given scenario through a one-on-one interview.
For example, to examine their skills as a manager, the assessor plays a role of a subordinate who is not satisfied with their current job.
Then, the assessee is asked to address the issue during the interview as the subordinate's manager.
To profoundly examine assessees' behaviors, assessors are required to act well in the given role and to strategically behave on the spot to simulate plausible situations that are difficult for the assessees.
This interview phase usually lasts for approximately 10--15 minutes and is video-recorded so that the assessors can review it later in the review phase.

In the review phase, the assessors play back the recorded video of the interview phase for each assessee.
This review phase usually lasts for 30 minutes.
Here, they inspect the assessee's behaviors, both verbal and nonverbal, and try to find cues for evaluating them that they might have missed during the interview phase.
Then, they make a decision on the assessee's skills and suitability for certain jobs (\eg ``This candidate is B+, suitable for being a manager to some degree, but has some room for improvement.'')
Furthermore, this review phase sometimes involves other assessors who independently check the video in order to validate the final decision.
The two assessors who attended this workshop often do it since they are senior to other assessors.

\subsection{What support assessors need}
\label{sec:requirements-need}

Then, we asked them about their ideas about how computers can help them to facilitate their assessment process.
Interestingly, both of them agreed that the final decision should be made by humans, not by computers.
This is because they thought computers would be incapable of making accurate decisions due to the complex nature of human assessment, and thus, they would ignore the output of computers when they have different opinions.
In other words, such a system would likely result in nothing additive for their assessment.
Moreover, if the system outputs only the final decision, it will easily cause a critical problem when their client asks for the reasons behind it, which assessors will not be able to explain.

On the other hand, they expected computers to help them assure the objectivity of their decision making.
In particular, they mentioned that the review phase involving watching the video of sessions is time-consuming and mentally-demanding, which challenges them in maintaining stable observation of assessees' behaviors.
In this sense, they agreed that systems that can automatically enumerate cues relating to their decision making would be beneficial in terms of achieving reliable and fair evaluation.

\subsection{What cues will be helpful}
\label{sec:requirements-cues}

Next, in order to enable computers to detect such insightful cues from the recorded videos of the interview session, we sought to elucidate their characteristics.
The two assessors agreed that they mainly try to check assessees' nonverbal behaviors (\eg their body movements) in the review phase.
This is because nonverbal behaviors are relatively implicit compared to verbal information such that their implications can sometimes be missed during the interview phase.
In response to this finding, we introduced some of the recent works in human behavior analysis to them and explained that such behaviors can be captured by computers to some extent.
We first showed demo videos of several computer vision techniques that digitize our behaviors, such as body pose estimation or facial keypoint detection \cite{DBLP:conf/iccv/FangXTL17}.
Then, we introduced works that estimate human behavioral features based on digitized behaviors, such as attention estimation based on human head poses \cite{arakawa2021mindless}, nod detection based on facial keypoints \cite{DBLP:conf/tvx/MaedaAR22}.
In addition, to provide better images of what computers can do for human assessment, we introduced a work by Sanchez-Cortes et al. \cite{DBLP:journals/tmm/Sanchez-CortesAMG12}, in which they utilized digitized behaviors to detect emergent leaders in a group discussion.
Their work introduced several hand-crafted features, such as the number of segments in which the amount of one's body movement exceeds a certain threshold, and applied a rule-based method using the features to calculate scores for each candidate.

Although we had anticipated that such works (\ie scoring based on some rules using human behavioral features such as nodding) could be helpful in human assessment as well, the two assessors both disagreed with it after contemplation, mentioning several reasons.
First, they clarified that their assessment process does not involve explicit counting of such behavioral features, nor do they believe it is meaningful.
They explained this reason with a simplified example; people who often nod are not necessarily suitable for managers.
Rather, they try to base their impression and judgment, which originally come from the interview phase, on those objective signals within context.
At this point, they mentioned that they often focus on scenes where the assessees showed unseen behaviors, such as sudden use of big gestures, because inferring what caused such changes could provide them insights that form the assessors' impression and judgment.

Secondly, the assessors were skeptical about the accuracy of the captured behavioral features (\eg attention level, nodding).
In detail, although they found the digitized behaviors shown in the demo videos precise to some extent, they questioned the validity of heuristics subsequently used to estimate those features (\eg time window and angular threshold of the head pose used to detect nodding).
Moreover, they were concerned about individual differences; such heuristics must be dependent on each assessee in order to precisely capture the features.
In sum, it would be hard for the assessors to trust a system based on such behavioral features (\eg the number of noddings) due to the limited validity of the algorithms used to estimate them.
Then, the system would result in being excluded from their process of decision making \cite{DBLP:conf/iui/YuBTZC19}.

\subsection{Our conclusion}
\label{sec:requirements-concl}

From the above discussion, we agreed that developing a system that outputs some indices about assessees based on behavioral features is not an optimal approach from the perspective of human-AI collaboration.
In contrast, we paid attention to the assessors' idea that extracting informative scenes based on assessees' nonverbal behaviors would be helpful to review the interview phase efficiently.
This approach represents the separation of observation and judgment; computers \textit{watch} the whole video on behalf of humans and humans \textit{make decisions} by checking scenes extracted by the computers.
Its benefit lies in the design that human assessors can make final decisions within the context of human-to-human communication, which is still hard for computers to deal with \cite{humanMachineTeaming}.
As a result, it would minimize the risk of losing the assessors' trust in the system due to uninterpretable decisions or unreliable indices made by the system.

In fact, the efficiency of such design (\ie separation of observation and judgment) in reflecting videos has been confirmed in the field of executive coaching \cite{DBLP:conf/chi/ArakawaY20}.
They applied unsupervised anomaly detection \cite{DBLP:journals/datamine/YamanishiTWM04} to time series data of digitized human behaviors (\eg body pose, facial keypoints) of coachees in coaching sessions.
Consequently, it was found that the interpretable presentation of scenes where coachees' nonverbal behaviors changed in comparison to other scenes enables professional coaches to reflect on the coaching sessions efficiently.
Since the algorithm to detect such scenes is based on unsupervised learning and does not involve pre-defined rules, it can mitigate biases that may arise in designing heuristics to estimate behavioral features (\eg attention and nodding).
Given the concern about such biases as the assessors mentioned in \secref{sec:requirements-cues}, we expected that utilizing unsupervised learning is crucially helpful when we seek to achieve human-AI collaboration in human assessment without losing assessors' trust.

To conclude, the assessors and the authors agreed to introduce the design of separating observation and judgment to facilitate their assessment in the review phase.
More specifically, we anticipated that computers would extract several scenes from the interview videos based on unsupervised anomaly detection and highlight them to help the assessors review the video.
We expected that maintaining the transparency of the system by employing such a design would prevent our project from falling into the dark side of introducing AI-based decision support tools \cite{doi:10.1080/0960085X.2021.1927213}.

\section{Study1: Algorithm Evaluation}
\label{sec:eval}

We then conducted a preliminary study to examine whether such an approach is feasible in terms of the algorithm's accuracy.
We applied the anomaly detection algorithm to assessment interview videos and examined how the extracted scenes were actually informative to assessors.
To this aim, we prepared 20 videos of the assessment interviews that had been conducted at the company before this study.
Neither of the two assessors participated in the interview sessions, and they were asked to assume the situation that they would be reviewing the videos independently, as we mentioned in \secref{sec:requirements-how}.
Thus, they reviewed the videos by focusing on verbal and nonverbal behaviors to find cues for evaluating assessees, as usual.
These assessment interviews were conducted online due to the COVID-19 situation, and each video was approximately 10-minutes long.

\subsection{Algorithm Detail}
\label{sec:eval-algorithm}

The algorithm we used is adopted from \cite{DBLP:conf/chi/ArakawaY19}, which is unsupervised anomaly detection with multimodal signals as input, as mentioned in \secref{sec:requirements-concl}.
Consulting with the assessors, we chose the following four nonverbal behavioral data as input to be used: facial keypoints, body pose, head pose, and gaze.
As mentioned in \secref{sec:requirements-cues}, each modality is accurately digitized from videos using recent computer vision techniques.
We used AlphaPose \cite{DBLP:conf/iccv/FangXTL17} to obtain facial keypoints and body pose.
Based on the estimated facial keypoints, the head pose was calculated by solving a perspective-n-point problem.
Finally, based on the facial keypoints and head pose, the gaze was estimated based on the RTGENE model \cite{DBLP:conf/eccv/FischerCD18}.

These behavioral data are calculated in every frame, and then the anomaly detection model processes its time-series data.
Specifically, the Gaussian Mixture Model (GMM) is fitted to the distribution of the frames in an online unsupervised learning manner.
Then, whenever new data come, the model outputs its outlierness based on the parameters obtained using the previous data while updating the parameters to fit the distribution including the new data.
In practice, the model processes the data window by window (\ie data observed within a specific duration) and outputs the outlierness on a batch basis.
Based on the sequence of the outlierness, we can identify anomaly scenes in the interview video that are likely to be informative to assessors as nonverbal cues.
In addition, the model is also capable of identifying the most anomalous and representative frame within each window.

Moreover, in this study, we extended their algorithm to enhance the interpretability of its output.
That is, we enabled the GMM model to output how each behavioral modality contributed to the estimated outlierness of the window.
This contribution of each modality is calculated as the change in the likelihood when we overwrite the parameters of the GMM model to ignore the corresponding feature in the estimation of the outlierness.
If the outlierness decreased largely, it implies that the cause of the outlier was likely due to the ignored feature, and vice versa.
In this way, we can identify the modality that contributed the most to the change by comparing the decrease of the outlierness.
We anticipated that this extension could provide assessors with further capabilities to interpret the model output.
This would meet the common practice in constructing better AI systems that recommends providing the clear attribution of their outputs to the corresponding inputs \cite{DBLP:conf/chi/AmershiWVFNCSIB19}.

\subsection{Procedure}
\label{sec:eval-procedure}

This study involved the two professional assessors who participated in the workshop \secref{sec:requirements}.
We asked them to independently review the 20 videos we prepared by focusing on the nonverbal behaviors of the assessees.
Then, they were requested to list the top 10 scenes for each video that are important to assess the assessee.
When they extracted such a scene, they were also asked to describe which behavior modality they based their thoughts on in a text, \eg ``At this moment, the assessee is making up his smile a little too much [facial keypoints]'' and ``Her eyes are scurrying and she is restless [gaze].''
Finally, we asked them to write evaluations on the assessees as they usually do.

\begin{figure}[tb]
    \centering
    \includegraphics[width=\linewidth]{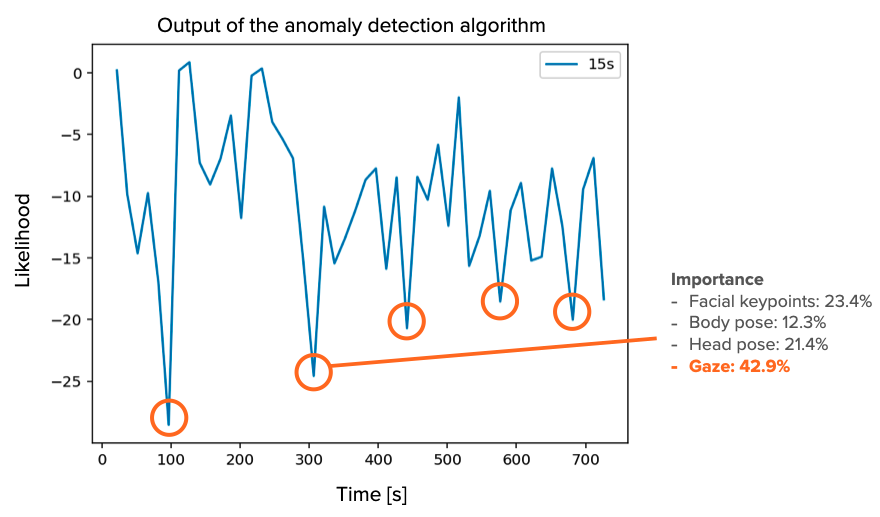}
    \caption{Example output of the anomaly detection algorithm applied to one interview video. It calculates time-series likelihoods of each scene (15-seconds window) based on multimodal behavioral data. For each detected scene, the algorithm outputs the importance of each modality. For example, the detected scene around 300 seconds most likely comes from the gaze data.}
    \Description{This figure shows a line chart illustrating the transition of the likelihoods over approximately 720 seconds. The line chart has five peaks indicating low-likelihood moments, which are highlighted using orange circles. One of them is annotated with an overlaying text, stating that the gaze information is the most important factor for the moment; its importance score is 42.9\%.}
    \label{fig:example-result}
\end{figure}

At the same time, the algorithm described in \secref{sec:eval-algorithm} processed each of the 20 videos.
The window size was set to 15 seconds.
An example of the algorithm's output (\ie time-series likelihood of scenes) is shown in \figref{fig:example-result}.
Based on the likelihood values, the algorithm then extracted the top 10 anomalous scenes according to the order of magnitude of the likelihood.
Finally, we compared the scenes extracted by the algorithm with those extracted by the two professional assessors.

\subsection{Findings}
\label{sec:eval-findings}

\subsubsection{Assessors sometimes use different cues for making decisions}
\label{sec:eval-findings-assessors}

Before examining the output of the algorithm, we first inspected the annotation provided by the professional assessors.
Interestingly, we found that the cues each assessor regarded as important diverged from those of each other to some extent.
Specifically, the proportion of cues that both assessors listed in the annotation data was 52.5\% among all the cues listed, while their overall evaluations for the 20 assessees were consistent.
From this result, we can infer that, even though the scenes the assessors had focused on were varied, they had something common in what they read from the cues, resulting in consistent evaluations.
It also implies that detecting such cues would be intractable for supervised learning because we will have difficulty in constructing clear criteria about which cues should be detected, hindering the preparation of training data.

This finding guided us to reframe the role of AI systems in human assessment: \textit{suggestion} rather than \textit{replication}.
In detail, considering that each assessor focuses on different cues in videos, we found that it is neither meaningful nor practical to detect cues that are completely coincident with their annotation.
Rather, given the review phase is prepared to find possibly missed cues (See \secref{sec:requirements-how}), suggesting possible cues that do not match the annotation data but are actually informative can be valuable.

\subsubsection{Interpretable anomaly detection provides effective and trustful suggestions}
\label{sec:eval-findings-benefit}

To dig into the above possibility further, we evaluated the output of the anomaly detection algorithm.
We first examined the agreement between the outputs and the annotation data.
As a result, we confirmed that approximately 38.0\% of the cues the assessors regarded as important had been detected as anomalous by our algorithm.
In addition, when we set the algorithm to enumerate the top 15 anomalous scenes, we found that the value of recall increased to 51.0\%, which was almost equivalent to the agreement rate between the two assessors.
This result supports the comments of the assessors (\secref{sec:requirements-cues}) and the conclusion we reached in the workshop (\secref{sec:requirements-concl}); scenes detected by applying unsupervised anomaly detection to digitized behaviors can serve as a basis for assessors' decision making.

\vspace{0.5mm}
\textbf{Cases of true-positive detection:}
We also qualitatively examined the detected cues with the guidance of the assessors regarding whether or not they are actually informative.
We first found that our algorithm worked well to detect important scenes even though the modality on which the algorithm focused can differ from that of assessors.
For example, the algorithm detected a scene that one of the professional assessors marked as important (\ie true positive), in which an assessee froze for a moment to figure out the best response to an assessor's critical question.
In this case, the algorithm displayed that the facial keypoints contributed the most to the outlierness of the scene.
This allowed us to infer that the algorithm detected the change in the keypoints around the assessee's mouse, which was caused by stopping the utterance, and the change in the keypoints around the assessee's eyes, which was caused by shutting the eyes tightly to ponder.

\vspace{0.5mm}
\textbf{Cases of false-positive detection:}
This room for interpretation provided by the algorithm was further beneficial when it detected a scene that was originally not marked as important by the professional assessors (\ie false positive).
For example, the algorithm detected a scene in which an assessee was emphasizing the words that had been repeatedly used to tell their dissenting opinion.
Here, the algorithm pointed out the change in the assessee's facial keypoints; indeed, the scene illustrated the figure of the assessee getting so close to the camera to emphasize their words that half of their keypoints disappeared.
Interestingly, the overall evaluation by the professional assessors for the assessee coincided with the detected scene.
Specifically, they remarked that the assessee ``was highly persistent'' and ``showed impatience, especially during the first half.''
After reflecting on this specific scene with the assessor, we agreed that this scene also has informative cues for the final judgment, which the assessors had not noticed in the review phase.
Given this case, we also agreed that, if the algorithm had not been designed to output such interpretable results, we might have ignored it because of our incapability of explaining its reason.
In other words, we would probably assume that such output is a negligible false-positive case of a black-box system.

Furthermore, we found that the algorithm yielded some (actual) false-positive results, such as a scene in which an assessee stopped their words for a moment to hold in a burp.
The scene was detected from the changes in the facial keypoints of the assessee but is a physiological phenomenon that is apparently not informative for the assessment.
Still, the transparent mechanism of the algorithm allowed the assessors to infer the reason behind the detection, which prevents the deterioration of their trust in the algorithm.

\subsubsection{Summary}

In sum, the performance of the algorithm was favorably received by the professional assessors.
In particular, they positively perceived that the design of separating observation and judgment, which we reached a consensus on in \secref{sec:requirements-concl}, would facilitate their assessment, although the algorithm does not completely replicate their annotation.
We inferred that there were two factors that made the assessors lean toward the acceptance of the design.
First, the design delegates a limited role (\ie observation) to the computer by considering its capability and allows human assessors to interpret various factors specific to each assessee that are highly human-contextual and difficult to capture by computers.
Second, the algorithm guides the assessors to infer the reason behind the detection by informing which of the nonverbal features contributes to the outlierness, which maintains their trust and prevents the ignorance of the outputs by the assessors.

\section{Study2: System Evaluation}
\label{sec:usability}

So far, we have identified system requirements and confirmed the feasibility of employing the unsupervised anomaly detection algorithm in human assessment.
Since our goal is to support assessors' decision making, we developed a dedicated interface incorporating the algorithm and evaluated its effectiveness and usability.

\subsection{Prototype}
\label{sec:usability-prototype}

\begin{figure}[tb]
    \centering
    \includegraphics[width=\linewidth]{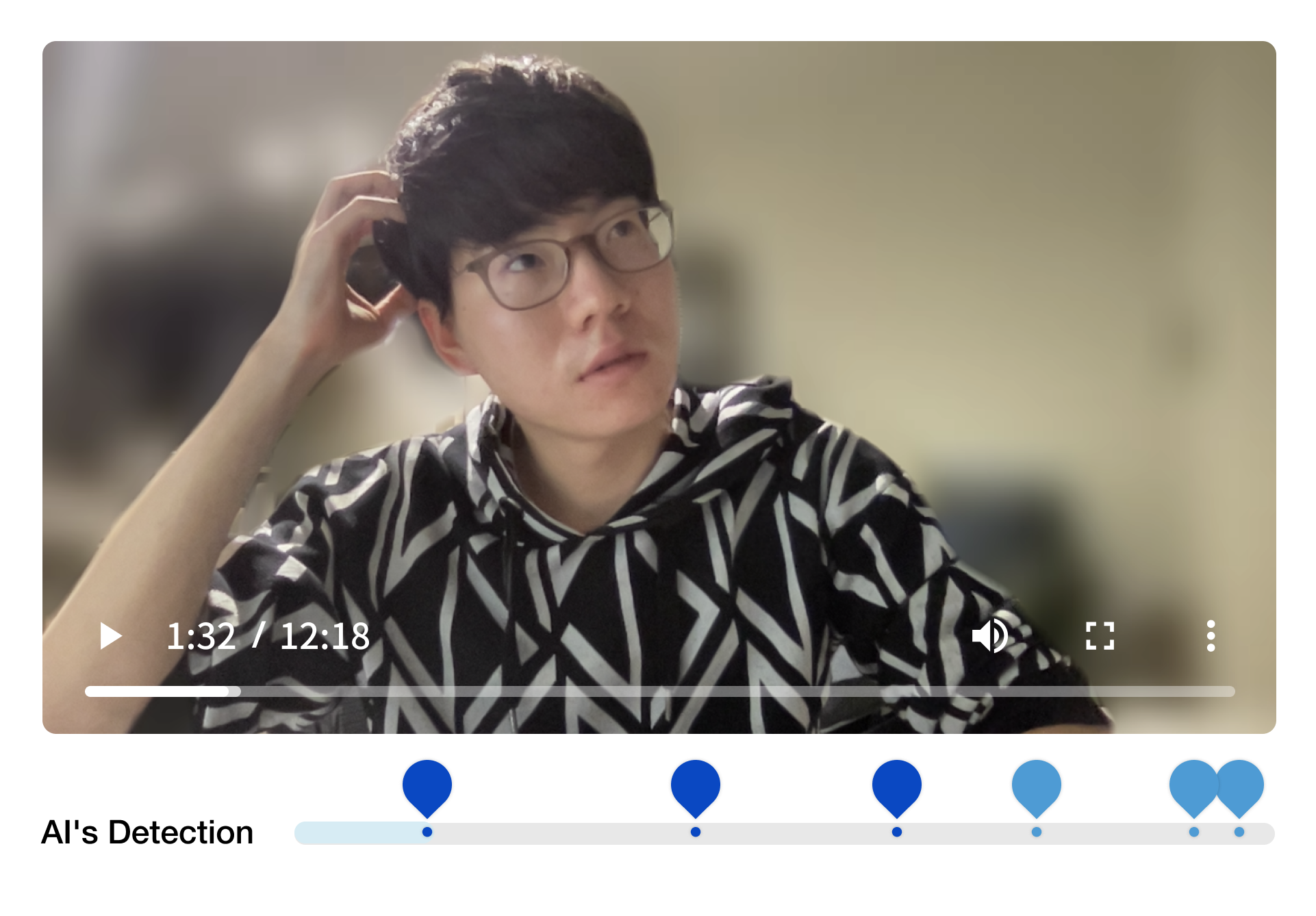}
    \caption{Prototype interface for supporting assessors' decision making used in Study 2.}
    \Description{This figure shows a playback interface for videos of a remote assessment interview. Below the video, there is a time-series sequence bar, on which pins indicating the detected anomaly points are placed. There are six pins on the bar; half of them are dark blue and the others are light blue.}
    \label{fig:prototype}
\end{figure}

\figref{fig:prototype} shows the interface we developed.
An interview video is embedded in its center part, along with a sequence bar on which pins indicating the detected anomaly points are placed.
We showed pins corresponding to the top six scenes based on the calculated outlierness.
The darker pins indicate scenes with top three outlierness while the lighter pins indicate scenes from the fourth to sixth.
When one of the pins is mouseovered, a popup is shown to describe the details of the scene, such as the modality that contributed to the detection.
When it is clicked, the assessor can jump to the corresponding scene within the video.
We implemented this interface as a browser-based application.

\subsection{Procedure}
\label{sec:usability-procedure}

We involved six professional assessors who had not participated in the previous two studies.
Two of them were junior assessors with experience of fewer than five years, while others were senior assessors with experience of more than ten years.
Similarly to Study 1, each assessor was asked to review four interview videos randomly chosen from the videos we used in Study 1.
However, they were also provided with the prototype and asked to use it for the assessment.
Before starting the review, we briefly explained the functions of the prototype using another sample video.
After they finished reviewing all videos, we conducted a video-based semi-structured interview in which we asked questions regarding the usability of the interface and its expected roles in their actual workflow.

\subsection{Results}
\label{sec:usability-results}

Overall, all assessors showed a favorable attitude toward the prototype with a willingness to incorporate it into their workflow.
The implications we obtained from their comments are summarized below.

\subsubsection{Deepened quality of the assessment}
\label{sec:usability-results-quality}

Five assessors described that their process of the assessment was supported by the prototype in several ways.
In particular, they felt that the objectivity of their decision was enhanced with the support, as they mentioned that the prototype showed some cues to which they would not have paid attention without it (\ie false-positive).
They described that such suggestions provided them a moment to reflect on their evaluation and some of the cues were actually informative to their final decision.
They also mentioned that the interpretability of such suggestions was critical to understanding the AI's suggestion and deepening their thoughts, corroborating the finding in \secref{sec:eval-findings-benefit}.
This result indicated that our approach of separating observation and judgment was helpful for the assessors.

The second aspect is that they gained confidence when the cues they used matched with the AI's suggestions (\ie true-positive).
This point was stressed by the junior assessors, who sometimes do not have full confidence in their decisions.
They appreciated that they could base their subjective feeling on AI's outputs that are objectively quantified through unsupervised learning.
After confirming the validity of taking the cues into consideration, it became easier for them to articulate their thoughts for the assessment.

Interestingly, they did not lose confidence when the cues they used were not in the AI's suggestions (\ie false-negative).
They understood that human assessment is a difficult task for AIs with many human contexts, and therefore, it is natural that AIs cannot detect some cues.
One of the junior assessors mentioned that they rethought such cases but could easily resolve the conflict by referring to other signals such as verbal information.
In other words, the assessor understood the mechanism and role of the AI (\ie extracting anomaly scenes based on nonverbal signals) and successfully used it complementarily for their decision.
This finding coincides with the previous discussion that assuring the interpretability of the AI and clarifying its boundary is a key to achieving a trustful human-AI collaboration \cite{DBLP:conf/chi/AmershiWVFNCSIB19}.

\subsubsection{Room for improvement of the prototype}
\label{sec:usability-results-improvement}

At the same time, the assessors gave several suggestions for the prototype.
The first is that they did not find the order of the outlierness helpful.
As we found in \secref{sec:eval-findings-assessors}, each assessor uses different cues for their decision.
As a result, the order of the importance of the detected scenes did not necessarily correlate with their evaluation of the importance of the scenes, leading to their confusion.
Still, they mentioned that they could easily resolve the conflict, similarly to the false-negative case we described above, \ie by acknowledging the capability of the AI and using other sources of information, such as verbal signals.

In addition, four assessors mentioned a need for incorporating paralanguage signals such as speech volume.
For example, a sudden change in speech volume often represents a cue when the person became passionate about the topic, an informative indicator for assessing the skill as a manager.
It is a benefit of the multimodal anomaly detection algorithm we employed that it can be extended to combine different modalities, and we are currently working toward it.

\subsubsection{Potential use scenarios}
\label{sec:usability-results-potential}

Overall, the assessors agreed that the prototype can be integrated into their workflow of human assessment to support their observation.
In addition, three assessors mentioned the potential use of the prototype in the feedback session as well.
In the feedback session, the assessor and assessee conduct a one-on-one session to reflect on the interview together after the review phase.
They mentioned the difficulty of providing effective feedback, especially when the assessee has a strong opinion and tends to ignore the feedback given through one-way communication from the assessor.
In such cases, the detected anomaly scenes could be used as visual and objective evidence of the assessee's behavior, offering a discussion ground for them to interpret the signals, which then corroborates the objectivity of the feedback.
In fact, such a capability of offering neutral perspectives in conducting reflection is a known benefit of introducing AI in highly-contextual situations \cite{DBLP:conf/chi/ArakawaY20}.

\section{Lessons Learned}

In this paper, we presented our case study conducted with professional assessors.
We first elucidated what they need from computer systems to realize human-AI collaboration in human assessment, a domain entwined with highly human contexts.
Based on their demands drawn in the workshop, we then evaluated the feasibility of employing unsupervised anomaly detection that can detect informative nonverbal cues from interview videos without relying on any heuristics.
Specifically, we examined how reliable its outputs are in terms of accuracy by comparing them with the annotations done by two professional assessors.
As a result, our algorithm can detect not only scenes that the assessors had focused on for their assessment but also those that had been unnoticed but informative, while with the capability of presenting them in an interpretable manner.
Lastly, we conducted a usability evaluation of a developed prototype involving six assessors.
The results elucidated the effectiveness of the system qualitatively, as well as directions for further improvement and potential use scenarios.
Overall, the series of studies confirmed the efficacy of our design, namely, the separation of observation and interpretation, in developing a supporting system for human assessment.

If we had started developing an algorithm to capture informative cues in the paradigm of supervised classification (\eg detecting specific human behavioral features such as nodding), it would not have worked effectively in concert with professional assessors.
First, it was suggested that different assessors look at different cues while having the same assessment result (\secref{sec:eval-findings-assessors}).
This inconsistency and unclear boundaries of classes make it difficult to prepare a dataset to train a model.
Moreover, even if we could train such a model based on supervised learning, it would lack interpretability and validity in its output, hindering the assessors from constructing a mental model about the behavior of the AI model.
As we see in \secref{sec:requirements-cues}, this would result in the failure of establishing trust in human-AI relationships \cite{Bansal2019}, leading the assessors to ignore the output of the model due to false positives.
In contrast, our approach allowed assessors to benefit even from false-positive detections, as shown in \secref{sec:usability-results-quality}.
Given these points, we conclude that the separation of observation and interpretation made possible by unsupervised anomaly detection will be a promising approach to building human-AI collaboration, especially in highly contextual domains that inevitably require a human decision.

\section{Limitation and Future Work}

The above findings were obtained from studies with a single assessment company, which might introduce a bias in the assessors' responses given that they had taken the same training program in the company.
We acknowledge that the number of assessors who participated in these studies was comparatively small.
To further investigate the role of AI in human assessment, future work is desirable to be conducted by involving more assessors from different enterprises.

We also would like to quantitatively explore how the introduction of the AI-based system affected the final evaluation of the assessors.
Here, without a large number of assessors and assessees, it is difficult to quantify it because the assessment process involves various factors, such as the diverse background of assessees and assorted assessment scenarios.
Moreover, this exploration would inform us of the contextual factors of the evaluation, such as the prospective capabilities of assessees and their fitness to the values of the assessees' enterprises.
While the current system entrusts the consideration of such factors to assessors, investigating how it can further contribute to the assessors' decision making would be a promising direction.

Another aspect to investigate is the advantage of the system's design regarding its trust with assessors.
In particular, as in previous literature on trust in human-AI collaboration \cite{Bansal2019,DBLP:conf/iui/YuBTZC19}, we would like to observe how assessors behave (\eg override the output or exhibit overreliance) when the accuracy of the system is artificially manipulated.

\subsection*{Acknowledgment}
We would like to thank the support of LEAD CREATE Inc., a Japanese human-assessment company.
This work was supported in part by JST ACT-X Grant Number JPMJAX200R and JSPS KAKENHI Grant Numbers JP21J20353.

\bibliographystyle{ACM-Reference-Format}
\bibliography{paper}

\end{document}